\documentclass[twocolumn,secnumarabic,superscriptaddress,amssymb, nobibnotes, aps, prl]{revtex4-2} 
\usepackage{epsfig,graphicx,subfigure,hyperref}
\hypersetup{
    colorlinks=true,
    linkcolor=blue,
    filecolor=blue,      
    urlcolor=blue,
    citecolor=blue
    }

\usepackage{amsmath,ulem}
\usepackage{xcolor}
\newcommand\ie {{\it i.e. }}
\newcommand{\p}{\partial}

\begin{document}

\title{Cavity Quantum Hall Hydrodynamics}

\author{Gabriel Cardoso}
\thanks{These two authors contributed equally}
\affiliation{Tsung-Dao Lee Institute, Shanghai Jiao Tong University, Shanghai, 201210, China}
\author{Liu Yang}
\thanks{These two authors contributed equally}
\affiliation{Tsung-Dao Lee Institute, Shanghai Jiao Tong University, Shanghai, 201210, China}
\affiliation{School of Physics and Astronomy, Shanghai Jiao Tong University, Shanghai 200240, China}
\author{Thors Hans Hansson}
\email{hansson@fysik.su.se}
\affiliation{Department of Physics, Stockholm University, AlbaNova University Center, 106 91 Stockholm, Sweden}
\author{Qing-Dong Jiang}
\email{qingdong.jiang@sjtu.edu.cn}
\affiliation{Tsung-Dao Lee Institute,
Shanghai Jiao Tong University, Shanghai, 201210, China}
\affiliation{School of Physics and Astronomy, Shanghai Jiao Tong University, Shanghai 200240, China}
\affiliation{Shanghai Branch, Hefei National Laboratory, Shanghai 201315, China}

\begin{abstract}
Motivated by recent experiments, we study the coupling of quantum Hall (QH) hydrodynamics to quantum electrodynamics (QED) within a resonance cavity. In agreement with experimental observations, we find that the Hall conductivity remains unchanged. However, the coupling to the cavity induces a second-order quantum reactance effect, contributing distinctly to the longitudinal AC conductivity. This effect arises from the exchange of energy between the QH fluid and cavity photons. Beyond the topological response, we show that the cavity couples to collective excitations, resulting in a shift of the Kohn mode frequency. Our methods  are broadly applicable to both integral and fractional  QH liquids, and our results offer a universal  perspective on the protection of  topological properties against long-range interactions induced by electromagnetic cavity modes.
\end{abstract}

\maketitle

\begin{figure}
\centering
\includegraphics[width=0.95\linewidth]{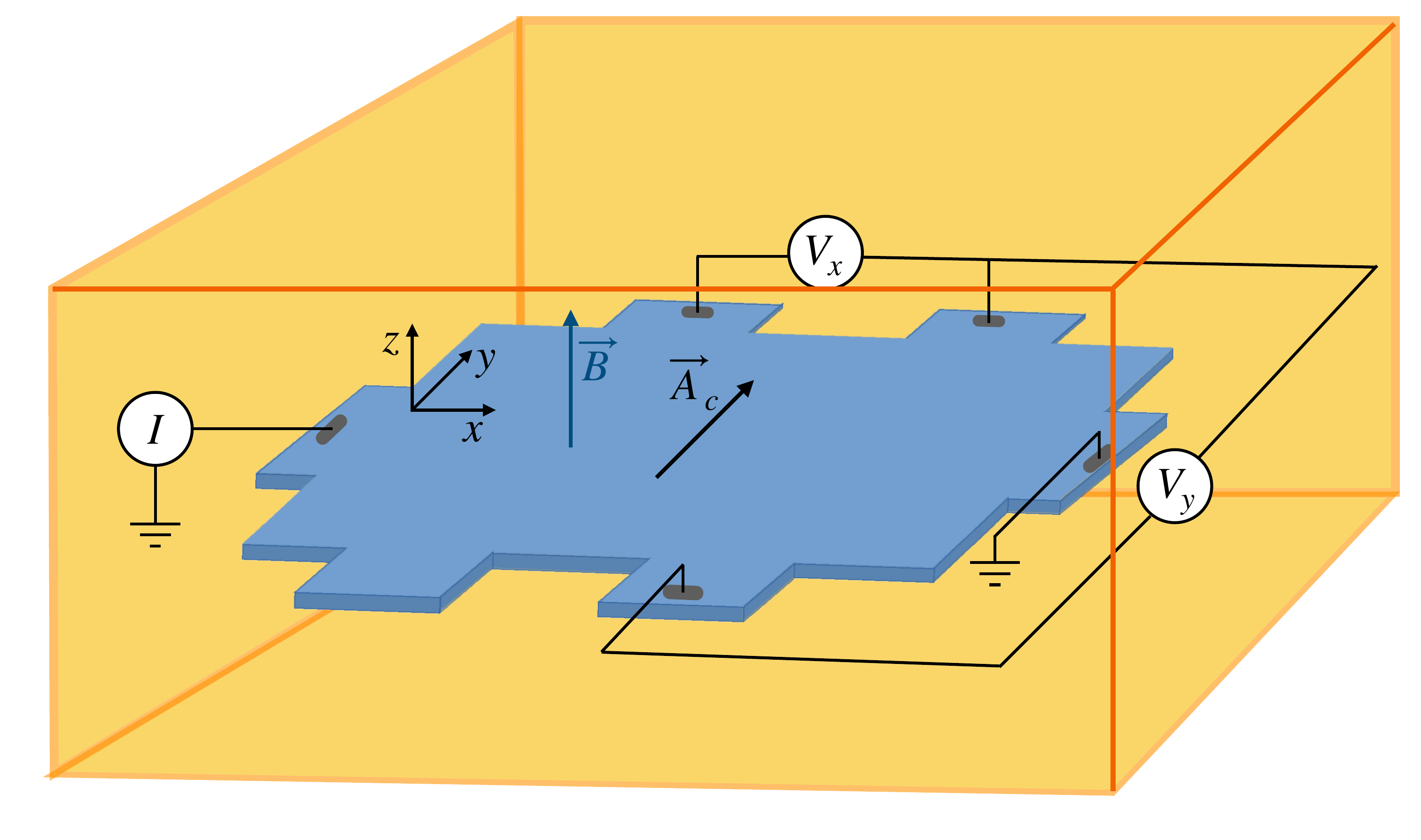}
\caption{Schematic view of an experiment to measure $\sigma^{xy}$ in a cavity. The Hall bar is in the $x$-$y$ plane, and we take the polarization of the lowest energy cavity mode as the $y$-direction. While the e.m. fluctuations in the cavity  do not modify $\sigma^{xy}$, they do modify the AC longitudinal conductivity in the $x$-direction, \ie transverse to the polarization of the dominant mode.}
\label{fig:Hallbar}
\end{figure}

\textit{Introduction}.—
Recent advances in optical cavities provide a new platform for investigating the response of matter to vacuum fluctuations of the electromagnetic field in both weak and strong coupling regimes \cite{ultrastrong2012,polariton2019,ultrastrong2019,ultrastrongRMP2019,RevQEDgas2021,Manipulating2021,chiral_cavity2021,review_cavity2022}. These developments have spurred numerous proposals for the cavity engineering of novel material properties, without the need for external pumping \cite{cavitygraphene2011,cavitySC2018,cavityMediated2019,cavitySC2020,cavityHall2021,amelio2021optical,cavityConductance2021,cavitySC2022,yaowang2023,bTMD2024,jiang2024flat,yang2024emergent,cavityTransition2023,wei2024cavity,Ciuti2024}. Within this context, important experiments have measured deviations from the quantized value of the  Hall conductance due to cavity fields \cite{2014Hallstate,cavityQHE2022,Ciuti2021,Rokaj2023}, with notable observations of the breakdown of topological protection at high filling factors. In sharp contrast, recent experiments at low filling factors have confirmed that there is no modification of the quantized Hall conductance \cite{Rokaj2024}. These findings raise fundamental questions regarding the robustness of topological material properties in the presence of fluctuating electromagnetic cavity fields, which can induce long-range interactions. In this work, 
we approach this issue from the perspective of quantum Hall (QH) hydrodynamics, specifically considering a Hall bar embedded within a cavity, as shown in Fig. \ref{fig:Hallbar}. The hydrodynamic framework applies in the regime where cavity photons do not cause inter-Landau level transitions, but is equally applicable to interacting and non-interacting systems. We demonstrate that while a linearly polarized cavity mode does not alter the Hall conductivity, it does induce a quantum reactance effect that significantly contributes to the AC longitudinal conductance, most strongly in the direction transverse to the cavity polarization. Furthermore, we show that the cavity mode couples to collective excitations of the QH liquid, resulting in a shift in the Kohn mode frequency.

The integer QH effect is observed when $n$ Landau levels are fully occupied, with the Hall conductivity $\sigma^{xy} = n\sigma_0 = ne^2/2\pi$ being measured to an extraordinary precision of less than $10^{-9}$ in sufficiently clean macroscopic samples at low temperatures \cite{quantumhall,1982Tsui,schopfer2013quantum,weis2011metrology}. The precise quantization is fundamentally rooted in electron band topology \cite{LaughlinArgument,Halperin1982,Laughlin2022}, and an integer invariant of the corresponding electronic wave functions \cite{TKNN1982,TNW}. In  strongly interacting systems,  electron correlations can lead to fractional values of the Hall conductance.
The first explanation of this remarkable phenomenon was in terms of  explicit model wave functions  for states in a partially filled Landau level, \cite{LaughlinArgument}, and later by mean-field theories based on composite bosons \cite{field1989,zhang1992chern,tournois2020}, and composite fermions \cite{jain2007composite,lopez1991fractional}.

In parallel with these microscopic approaches, there exists a well-developed hydrodynamic theory for QH liquids, which provides a unified description of the topological properties of both integer and fractional states \cite{zhang1992chern,wen1992shift}. This theory can be derived, for instance, from a theory of composite bosons, and, to the lowest order in a derivative expansion, it corresponds to the theory proposed by Wen and Zee \cite{[{For the development of the hydrodynamic theory of the QHE, see also  }]stone1990superfluid,*lee1991collective}. By including higher-order derivative terms, this theory can also describe non-topological effects, such as collective excitations \cite{lopez1991fractional}. The conventional framework describes the coupling of QH liquids to external electromagnetic fields but typically neglects the effects of electromagnetic field fluctuations. In this Letter, we extend this approach to investigate the universal impact of cavity electromagnetic fluctuations on QH liquids. We show that our universal hydrodynamic results are consistent with previous microscopic calculations that used filled Landau levels of non-interacting electrons to study cavity effects on integer QH states. Finally, we discuss various possibilities to experimentally verify our predictions.

\textit{Hydrodynamic description}.—We shall use the hydrodynamic action,
\begin{align}
    {\cal L}_{H} = -\frac m {4\pi} \epsilon^{\mu\nu\rho}b_\mu\p_\nu b_\rho-\frac e {2\pi} \epsilon^{\mu\nu\rho}A_\mu\p_\nu b_\rho \nonumber  \\ + \frac m {4\pi \omega_B }\vec E_b^2 - \frac u 2 B_b^2,\label{eq:hydrofull}
\end{align}
where $E_b^i = -\partial_t  b^i - \partial^i b^0$,  $B_b = \epsilon^{ij}\partial_i b_j$, and $j^\mu = \epsilon^{\mu\nu\rho}\partial_\nu b_\rho$ is the electric current which couples to the external electromagnetic field $A_\mu$. The first two terms give the topological Wen-Zee action for a $ \nu = 1/m$ QH liquid \cite{wen1992shift}, and the two last are higher derivative, non-topological terms \cite{wire2019}. Here, $\omega_B =eB/m_e$ is the cyclotron frequency, and the constant $u$ is the strength of  a repulsive delta function  potential. The generalization to an arbitrary potential $u(q)$ is straightforward \cite{zhang1992chern}. We use  units where $\hbar=c=\varepsilon_0=1$.

For a single linearly-polarized cavity mode, 
\begin{equation}
    \vec{A}_c(t,\vec{x})=q(t)\phi_{k_c}(\vec{x})\hat e_y,\label{eq:Acsinglemode}
\end{equation}
where $\hat e_y$ is a unit polarization vector (see Fig.~\ref{fig:Hallbar}), $\phi_{k_c}$ is the normalized mode function satisfying $\Delta\phi_{k_c} = -k_c^2\phi_{k_c}$, and $q(t)$ is the fluctuating amplitude. For this single mode, the usual Maxwell Lagrangian ${\cal L}_M = \frac\varepsilon{2}\vec E^2-\frac 1{2\mu}\vec B^2$ reduces to 
\begin{align}
    L_c=\frac \varepsilon{2} (\dot{q}^2-\omega_c^2q^2)\, ,\label{eq:maxwellcavity}
\end{align}
 where $\omega_c=k_c/\sqrt{\varepsilon\mu}$ is the cavity frequency. For a rectangular cavity, as shown in Fig.~\ref{fig:Hallbar}, $k_c$ is the momentum of the lowest TE mode. In relevant experimental settings, one expects the cavity effects on the QH liquid to be dominated by a single lowest energy mode~\cite{Rokaj2024}. We note that the  frequency and amplitude of this mode are determined both  by the cavity geometry, which specifies $k_c$, and by the effective values of the relative permittivity $\varepsilon$ and permeability $\mu$.

\textit{Topological response}.—We first calculate the topological response by dropping the higher-order terms in \eqref{eq:hydrofull},
\begin{align}  
   S =  &\int({\cal L}_{M}+{\cal L}_{H})= \int dt\Big[\frac{\varepsilon}{2}(\dot{q}^2-\omega_c^2q^2)\hfill   \label{eq:hydrofull2}    \\
    &\hfill-\int d^2x\left(\frac{m}{4\pi}\epsilon^{\mu\nu\rho}b_\mu\p_\nu b_\rho+\frac{e}{2\pi}\epsilon^{\mu\nu\rho}(A+A_c)_\mu\p_\nu b_\rho\right)\Big],\nonumber
\end{align}  
where $A_c$ is the projection of the (three-dimensional) cavity field \eqref{eq:Acsinglemode} on the (two-dimensional) plane of the Hall bar, and for simplicity we assume that the polarization vector of the cavity mode lies in this plane. Also, $A$ is a non-dynamical background probe field. With this the   the coupling term becomes
\begin{equation}
    -\frac e{2\pi}\int dt q(t)[\p^t b^{x}_{k_c}-(\p^x b^t)_{k_c}],
\end{equation}
where $f_{k_c}(t) = \int d^2x\phi_{k_c}(\vec x)|_H f(t,\vec x)$ denotes the average of the function $f(t,\vec x)$, weighted by the cavity mode function, evaluated on the Hall bar.

Since \eqref{eq:hydrofull2} is quadratic in both $b_\mu$ and $q$, we can integrate these to get the effective response action
\begin{align}
    &S_{\rm eff}[A]=-\frac{e^2}{4\pi m}\int\frac{d\omega}{2\pi}\Big\{\int d^2x\,i\omega\epsilon^{ij}A_iA_j\hfill\\
    &\hspace{.35\linewidth}+\frac{2}{\varepsilon}\left(\frac{e^2}{4\pi m}\right)\frac{\omega^2}{\omega^2-\omega_c^2}A_{x,k_c} A_{x,k_c}\Big\},\nonumber
\end{align}
where we simplified the expressions by using the radiation gauge, $A^0=\vec\nabla\cdot \vec A=0$, for the probe field. The corresponding electric current is 
\begin{align}
    j^k(\omega,\vec x) =& -\frac{1}{m}\frac{e^2}{2\pi}i\omega\epsilon^{kl}A_l(\omega, \vec x)     \label{eq:current}  \\
    &\hspace{.05\linewidth}+ \frac{1}{m}\frac{e^2}{2\pi}\frac{2\alpha}{m\varepsilon}\frac{\omega^2}{\omega^2-\omega_c^2}\delta^{k}_xA_{x,k_c}(\omega)\phi_{k_c}(\vec x).\nonumber 
\end{align}
We recognize the leading term as the Hall conductivity. The correction is parametrically suppressed by the ratio $\alpha/\varepsilon$, where $\alpha=\frac{e^2}{4\pi}$ is the fine-structure constant, and is spatially modulated by the cavity mode function. Moreover, only the component of the probe field which is transverse to the cavity polarization, $A_x$, gives an additional  current,  $\propto\delta^{k}_x$ , which is also transverse to the cavity polarization. It follows that only the longitudinal conductivity is modified.

Let us assume that the Hall bar is small compared to the cavity dimensions and approximate both $A$ and $A_c$ as homogeneous (long wavelength approximation). Then the the current \eqref{eq:current} simplifies to
\begin{align}
    j^x(\omega) &= \frac{e^2}{2\pi}\frac{1}{m}E_y(\omega) +\frac{e^2}{2\pi}\frac{1}{m^2}\frac{2\alpha}{ L_{\rm eff}}\frac{i\omega}{\omega^2-\omega_c^2}E_x(\omega),\label{eq:current2}\\
    j^y(\omega) &=-\frac{e^2}{2\pi}\frac{1}{m}E_x(\omega)\, ,\label{eq:jcavity}
\end{align}
where the length scale $L_{\rm eff}=\varepsilon/(\phi^2A_H)$ is proportional to the ratio of the quantization volume of the cavity $V_c$ to the area of the Hall bar $A_H$, since $\phi$ is normalized to unity. We believe that this simplified description captures the essence of the experiment in \cite{Rokaj2024}. We see that the Hall conductivity is not modified, in agreement with the experimental findings. Interestingly, we find an additional longitudinal contribution to the AC conductivity. It vanishes in the DC limit $\omega\to 0$, consistent with the quantum Hall phenomenology. Its magnitude is controlled by the ratio $\alpha/L_{\rm eff}$, which is related to the energy scale of the light-matter coupling. As expected this effect is enhanced for smaller cavities, where quantum fluctuations are stronger.

\begin{figure}[]
\centering
\includegraphics[width=0.95\linewidth]{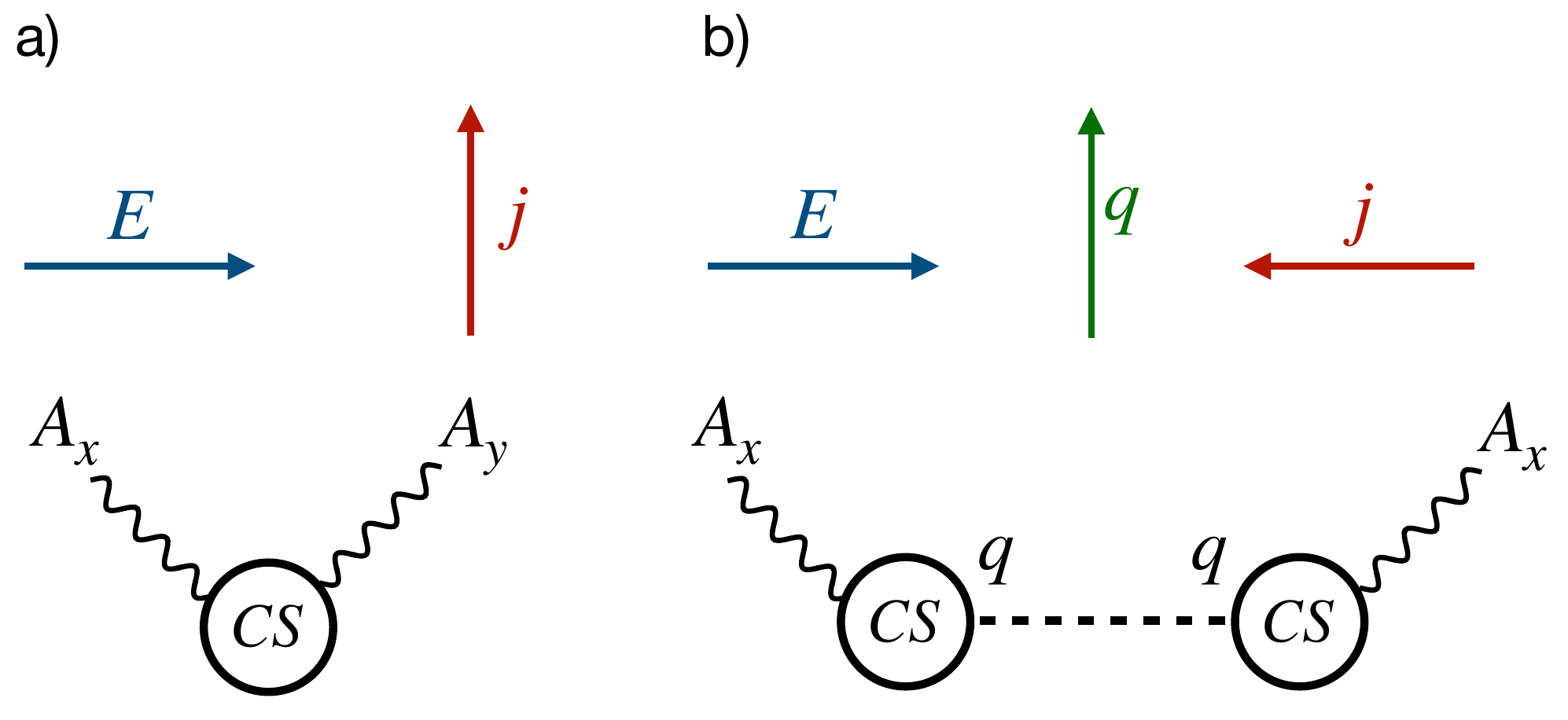}
\caption{Pictorial representation of the effective response action of the cavity-Hall system. a) The Chern-Simons two-point vertex, which gives a transverse current response to an applied field. b) The longitudinal term, which results from connecting two Chern-Simons vertices by a cavity propagator. This is a second-order effect where an applied field leads to the exchange of cavity photons with transverse polarization, which in turn results in a longitudinal current. The dissipationless periodic exchange of energy between the QH liquid and the cavity generates an effective quantum reactance.}
\label{fig:ACterm}
\end{figure}

\textit{Quantum reactance}.—To understand the nature of the longitudinal response $\sigma_{xx}$ in \eqref{eq:current2}, it is useful to recall the current response in an LC circuit,
\begin{equation}
    I(\omega) = -\frac{1}{L}\frac{i\omega}{\omega^2-\omega_0^2}V(\omega),
\end{equation}
where $\omega_0=1/\sqrt{LC}$ is the resonant frequency. The parallel to \eqref{eq:current2} is obvious, and in particular the current and voltage (or in our case, $j_x$ and $E_x$) are out of phase by an angle $\pi/2$. In the LC circuit energy is transferred without loss between the capacitor and the inductor, while in our case it is transferred between the electrons in the QH liquid and the cavity field. In neither case is there any dissipation due to the phase lag; the impedance is purely reactive. Note that for an ideal (\ie very high $Q$) cavity,  we find that the Hall-cavity setup will act as a perfectly dissipation-less band-pass filter at the cavity frequency. For a lossy cavity (\ie finite $Q$), the response will be that of an RLC circuit, with effective resistance proportional to the cavity loss (see our accompanying paper \cite{chiral2024}). The interplay between the QH liquid and the cavity field is illustrated in Fig.~\ref{fig:ACterm}, where panel a) shows the usual Hall effect: An applied field generates a current in the transverse direction. Panel  b) shows the following second-order process: First an applied field leads to the emission of a transverse cavity photon, then the absorption of a cavity photon generates a transverse current, so that finally the current response is parallel to the applied field. This is the quantum reactance effect. Note that this picture also explains why the effect is induced only in the direction transverse to the polarization of the cavity mode.

\textit{The Kohn mode}.—We now go beyond the topological response and  reintroduce the higher-order terms in \eqref{eq:hydrofull}. The resulting dynamics describes the Kohn mode at high frequency $\omega=\omega_B$. The effect of the cavity can be seen by taking a homogeneous $b_i(t)$. Going to the frequency domain and  using the radiation gauge,  the  effective action becomes,
\begin{align}
    &\int\frac{d\omega}{2\pi}\Big\{A_H\left[\frac{m}{4\pi}i\omega\epsilon^{ij}+\frac{m}{4\pi\omega_B}\omega^2\delta^{ij}\right]b_ib_j\nonumber\\
    &\hspace{.35\linewidth}-\frac{1}{2\varepsilon}\left(\frac{e}{2\pi}\right)^2\frac{\omega^2}{\omega^2-\omega_c^2}b_{k_c}^x b_{k_c}^x\Big\}\, .
\end{align}
The Euler-Lagrange equations have a solution corresponding to the Kohn mode, with the modified frequency
\begin{equation}
    \omega = \omega_B + \frac{1}{m}\frac{\alpha}{ L_{\rm eff}}.\label{eq:w_K}
\end{equation}
Thus the coupling to the cavity mode gives an $O(\alpha/ L_{\rm eff})$ correction to the Kohn mode frequency. Similarly, one can compute the modification of the conductivities due to the higher-order terms in the hydrodynamic action. The corrections are suppressed at small frequencies by powers of $(\omega/\omega_B)$, which vanish in the topological limit $\omega_B\to\infty$. From the microscopic point of view they arise from the transitions between Landau levels, as we now discuss (see also \cite{chiral2024}). 

\textit{Comparison with previous work}.—While the above derivation captures the universal aspects of the response of the QH liquid to the probe and cavity fields, it does not capture any details about the microscopics behind the phenomenology. 
For the integer effect, however, one can address this by coupling a free 2d electron gas filling an integer number of Landau levels to the cavity mode. This problem was studied in Ref.~\cite{Rokaj2023} by Rokaj et al. in terms of the Hamiltonian
\begin{align}
    H_\text{CM}&=\frac{(\vec{\pi}+e\sqrt{N}\vec{A}_c)^2}{2m_e}+\omega_c\left(a^\dagger a+\frac{1}{2}\right),\\
 \nonumber
\end{align}
for the (renormalized) center of mass coordinate of the electrons with an effective mass $m_e$, and the cavity mode $\vec{A}_c=A(a+a^\dagger)\hat{e}_y$, which is taken to be constant over the Hall sample.

Starting from the commutation relations $[\pi^x,\pi^y]=-ieB$ and  $[a,a^\dagger]=1$, one can diagonalize $H_\text{CM}$ in terms of new canonical raising and lowering operators $b^\dagger_\pm$, $b_\pm$ \cite{Rokaj2023},
\begin{align}
     H_\text{CM} =\sum_{j=\pm}\omega_j(b^\dagger_j b_j+\frac{1}{2})\, .\label{eq:diag_H}
\end{align}
The frequencies $\omega_\pm$ are given in the supplementary material in terms of $\omega_B$, and $ \widetilde{\omega}_\text{c}=\sqrt{\omega_\text{c}^2+\omega_d^2}$, where the diamagnetic frequency is $ \omega^2_d=e^2 N/({m_e \varepsilon V_c}) = {2\alpha\nu\omega_B}/{L_{\rm eff}}$ for the case of $\nu$ fully occupied Landau levels, corresponding to the center of an integer Hall plateau. We recognize the characteristic energy scale $\omega_d^2/\omega_B=2\nu\alpha/ L_{ \rm eff}$ of the light-matter coupling in the cavity. The higher frequency $\omega_+$ corresponds to the Kohn mode, while the lower frequency $\omega_-$ is close to the cavity mode. Expanding in $\omega_c/\omega_B$ we find, to leading order,
\begin{align}
    \omega_+=\omega_B+\frac{\alpha\nu}{ L_\text{eff}},
\end{align}
which for integer filling fractions $\nu$ ($1/m$) agrees with the value \eqref{eq:w_K} obtained from the hydrodynamics.

\begin{figure}[]
\centering
\includegraphics[width=0.98\linewidth]{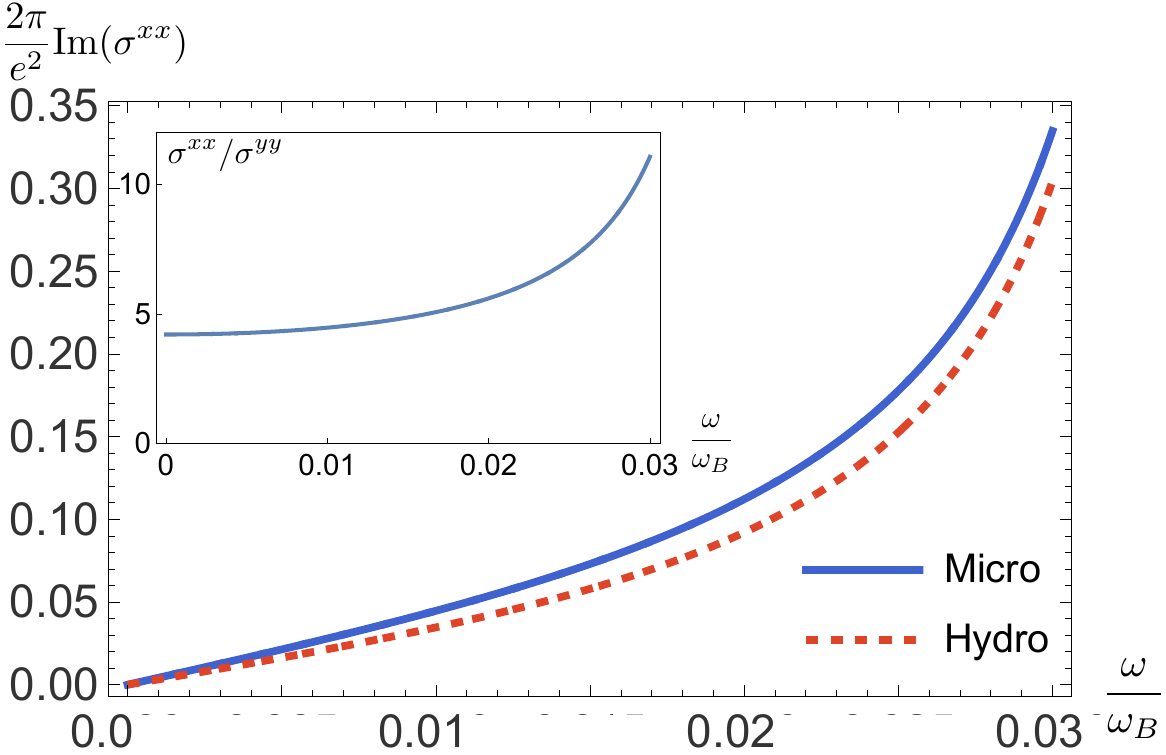}
\caption{Full AC conductivity $\sigma^{xx}(\omega)$ obtained from the microscopic calculation, Eqs.~(\ref{eq:sigmaxxmicro}) and (\ref{eq:sigmayymicro}) (Micro), and the quantum reactance contribution obtained from the hydrodynamic calculation (Hydro). The conductivity is shown in units of $e^2\nu/(2\pi)$ and the frequency is in units of the cyclotron frequency $\omega_B$. Here we consider a Hall bar in a cavity with the parameters from recent experiments \cite{cavityQHE2022,Rokaj2024}: a sample area of $A_H=40\times200\mu\text{m}^2$, a filling ratio of $\nu=1$, a cyclotron frequency of $\omega_B=24.3\text{THz}$, and a cavity frequency of $\omega_c=0.88\text{THz}$, while we set the effective length $L_\text{eff}=3.38\mu\text{m}$, corresponding to an effective cavity volume reduced to 10\% of the original experimental size $V_c=2.7\times10^5\mu\text{m}^3$. We see that for these parameters the quantum reactance contribution from Hydro dominates $\sigma^{xx}$. The inset shows that the ratio  $\sigma^{xx}/\sigma^{yy}$ from Micro is large, which demonstrates the anisotropic effect from the $y$-polarized cavity mode. Note that by the Hydro prediction, this ratio would be infinite as $\sigma^{yy}$ is zero.
}
\label{fig:plotsigmaXX}
\end{figure}

For a linearly polarized cavity,  Ref.~\cite{Rokaj2023} gives the  results for the AC conductivities (or, more precisely, the conductances, although we shall not make this distinction). For $\nu=1/m=1$, we can now  compare these with the results \eqref{eq:current2} and \eqref{eq:jcavity}, obtained from the topological part of the hydrodynamic action, in the parameter regime  $\omega_B \gg\widetilde{\omega}_c>\omega_c>\omega_d$. Expanding the results from Ref.~\cite{Rokaj2023}  to second order in $\omega_d/\omega_B$, we find 
\begin{align}
     \frac{\sigma^{xx}(\omega)}{e^2\nu/(2\pi)}&=\frac{i \omega\,\omega_B}{\omega_B^2-\omega^2}+\frac{2\alpha\nu}{L_\text{eff}}\frac{i \omega}{(1-\frac{\omega^2}{\omega_B^2})^2(\omega_c^2-\omega^2)}\, ,\label{eq:sigmaxxmicro}\\
    \frac{\sigma^{yy}(\omega)}{e^2\nu/(2\pi)}&=\frac{i \omega\,\omega_B}{\omega_B^2-\omega^2}+\frac{2\alpha\nu}{L_\text{eff}}\frac{i \omega_B^{-2}\omega^3}{ (1-\frac{\omega^2}{\omega_B^2})^2(\omega_c^2-\omega^2)}\,,\label{eq:sigmayymicro}\\
    \frac{\sigma^{xy}(\omega)}{e^2\nu/(2\pi)}&=\frac{\omega_B^2}{\omega_B^2-\omega^2}+\frac{2\alpha \nu}{ L_\text{eff}}\frac{\omega_B^{-1} \omega^2}{(1-\frac{\omega^2}{\omega_B^2})^2(\omega_c^2-\omega^2)}\label{eq:sigmaxymicro}\, .
\end{align}
In the topological limit $\omega/\omega_B\to 0$, the only modification due to the cavity is to the $\sigma^{xx}$  in \eqref{eq:sigmaxxmicro}, which agrees with the quantum reactance term in \eqref{eq:current2}. However, for a finite $\omega/\omega_B$, the remaining terms with poles at $\pm\omega_B$ account for transitions between the Landau levels. At low frequencies they give a linear contribution to both $\sigma^{xx}$ and $\sigma^{yy}$. 
For parameters in the relevant experiments~\cite{cavityQHE2022,Rokaj2024}, these terms can be comparable to the quantum reactance term (see the supplemental material).  For smaller cavities, the quantum reactance term can dominate the full longitudinal AC conductivity (see Fig.~\ref{fig:plotsigmaXX}). 
For the Hall conductivity, the leading correction appears only in quadratic order in $\omega$, and it vanishes in the topological limit $\omega/\omega_B\to 0$. 

In the zero-frequency limit, $\sigma^{xx}=\sigma^{yy}\to0$, and it is instructive to express the Hall conductivity as
\begin{equation}  \label{eq:formula}
       \sigma^{xy}(\omega)=\frac{n_e}{B^2}\langle\mathbf{0}|i[\pi^x,\pi^y]|\mathbf{0}\rangle + O(\omega)\, ,
\end{equation}
where $n_e$ is the electron density and $|\mathbf{0}\rangle$ is the ground state of $H_{\text{CM}}$. For the fully occupied Landau levels with filling ratio $\nu$, this gives the quantized DC Hall conductivity as $ e^2\nu/(2\pi)$, independent of the polarization of the cavity mode (see supplementary material). Eq. \eqref{eq:formula} also illustrates an important point about the calculations based on using the CM degree of freedom only, as it is precisely the expression for the response of a single charge $Ne$ particle moving in a magnetic field, and has no relation to topology. It follows directly from a classical argument based on Galilean invariance~\cite{Kohntheorem}, and cannot explain the presence of a plateau \cite{prangebook}. The hydrodynamic theory, on the other hand, incorporates the topological properties of both the integer and the fractional quantum Hall states \cite{[{For a recent microscopic study, see }]bacciconi2024theory}, and can be used to calculate the finite-momentum response in the case of inhomogeneous fields.

\textit{Conclusions and outlook}.—Our results give a new perspective on  recent experimental and theoretical efforts to understand the effect of cavity fluctuations on quantum Hall liquids in the zero temperature limit. In the $\nu=1$ the experiment reported in \cite{Rokaj2024}, $\omega_B=24.3\,\text{THz}$, which is much greater than the cavity frequency $\omega_c=0.88\,\text{THz}$, which is again greater than the diamagnetic frequency $\omega_d=0.50\,\text{THz}$ for a $40\times200\mu\text{m}^2$ Hall bar. The experiment was performed at temperatures lower than 50mK, where thermal fluctuations are smaller than 6.5 GHz, so finite-temperature effects can be neglected. Moreover, the spectrum of the cavity is such that the single-mode approximation is applicable, with the other modes lying at much higher energies. As we have shown in this work, this setup can be studied in terms of the quantum Hall hydrodynamics coupled to a single linearly polarized cavity mode. 

We found that the cavity mode gives a contribution to the longitudinal conductivity in the direction transverse to the polarization through a second-order quantum reactance effect. This effect is mediated by the dissipation-less exchange of energy between the QH liquid and the cavity photons which, to our knowledge, has not been reported before. We note that, whilst this effect reaches resonance at the cavity frequency, which is typically in the THz range, the effective impedance can still be measured at small probe frequencies, in which case it is purely reactive for an ideal cavity. The resonance can in principle be used as a band-pass filter which selects the cavity frequency. For a cavity formed by movable mirrors, the output frequency can be tuned. We compared our findings with previous calculations for the integer case in terms of the center-of-mass Hamiltonian \cite{Rokaj2023}. We also showed that this approach predicts that corrections to the Hall conductance at the center of the plateau are absent for an arbitrary cavity polarization, at least in the case of a homogeneous cavity field. Besides, going beyond this limitation, the hydrodynamic formalism also by construction incorporates all topological aspects of the Hall response.

The work in this paper can be extended in a number of interesting directions.  The hydrodynamic calculation of the AC conductivities can be extended to higher frequencies by including higher-order contributions such as the Maxwell terms in \eqref{eq:hydrofull}, and terms related to the Hall viscocity~\cite{gromov2014}. Another extension is to finite temperature by using the Keyldish formalism, and to lossy cavities by having a finite $Q$ number. Chiral cavities, that support only a single circularly polarized mode, could introduce time-reversal symmetry breaking fluctuations, potentially giving rise to novel phenomena as suggested by previous studies in chiral quantum electrodynamics{ \cite{PhysRevB.99.201104,PhysRevB.104.L081408,chiral_cavity2021,chiralQED2022,PhysRevX.13.031002}. }Several of these ideas will be explored in a future paper \cite{chiral2024}, where we also investigate the particle nature of the charge carriers in the coupled system. 

\textit{Acknowledgements}.—
THH thanks the director and members of the T.D. Lee Institute for kind hospitality during October of 2024. We gratefully acknowledge previous helpful discussions and feedback from Qian Niu, Kun Yang, and Alexander Abanov. This work was supported by National Natural Science Foundation of China (NSFC) under Grant No. 12374332,  Jiaoda 2030 program WH510363001-1, the Innovation Program for Quantum Science and Technology Grant No. 2021ZD0301900. 

\bibliographystyle{apsrev4-1}
\bibliography{apssamp}

\newpage
\onecolumngrid
\appendix 
\clearpage
\renewcommand\thefigure{S\arabic{figure}}    
\setcounter{figure}{0} 
\renewcommand{\theequation}{S\arabic{equation}}
\setcounter{equation}{0}
\renewcommand{\thesubsection}{SM\arabic{subsection}}

\begin{center}
\textbf{\large Supplemental Material: Cavity Quantum Hall Hydrodynamics}
\end{center}

\section{I. Mode functions on the Hall bar}

The mode function is defined in the three-dimensional cavity region, with
\begin{align}
    &\Delta\phi_{k_c} = -k_c^2\phi_{k_c}, \quad\quad \int_{V_c} d^3x|\phi_{k_c}(\vec{x})|^2=1,\label{eqsup:psi}
\end{align}
which in particular includes a normalization with respect to the cavity volume. A useful approximation is to assume that the mode function is constant on the Hall bar, which we take to lie on the $z=z_H$ plane, so that
\begin{equation}
    \phi_{k_c}|_H(\vec x) = \phi_{k_c}(x,y,z_H)\approx \frac{\chi_c}{\sqrt{V_c}},
\end{equation}
with $\chi_c$ a dimensionless constant determined by the geometry. A particular formula which often appears is the averaging over the cavity mode function,
\begin{equation}
    \int_{A_H} d^2x\,\phi_{k_c}|_H(\vec x) = \frac{\chi_c A_H}{\sqrt{V_c}}.
\end{equation}
This formula is true more generally. For example, for a box geometry,
\begin{align}
    &\phi_{k_c}(\vec x)=\frac{2\sqrt{2}}{\sqrt{L_xL_yL_z}}\sin\left(\frac{\pi x}{L_x}\right)\sin\left(\frac{\pi y}{L_y}\right)\sin\left(\frac{\pi z}{L_z}\right), &k_c = \pi\sqrt{\frac{1}{L_x^2}+\frac{1}{L_y^2}+\frac{1}{L_z^2}},
\end{align}
and assuming that the Hall bar extends over the whole $A_H=L_xL_y$ area on the $z=L_z/2$ plane, one finds
\begin{equation}
    \chi_c = \frac{\sqrt{V_c}}{A_H}\int d^2x\,\phi_{k_c}(x,y,L_z/2) = \frac{8\sqrt{2}}{\pi^2}\approx 1.15.
\end{equation}
Alternatively, if the Hall bar occupies a small enough region inside the cavity so that the mode function can be approximated as nearly constant, one often finds the combination
\begin{equation}
    \phi_{k_c}|_H(\vec x)\int_{A_H} d^2x\,\phi_{k_c}|_H(\vec x) = \frac{\chi_c^2 A_H}{V_c}\equiv\frac{\varepsilon}{L_{\rm eff}},
\end{equation}
where we define a characteristic length scale $L_{\rm eff}$ of the cavity-Hall system geometry for the later convenience.

\section{II. Kohn mode}

Integrating out the cavity fluctuations, we obtain the effective action
\begin{align}
    &\int\,dt\,d^2x\left[-\frac m {4\pi} b\,db + \frac m {4\pi \omega_B }\vec E_b^2 - \frac v 2 B_b^2\right]+\frac{1}{2\varepsilon}\left(\frac{e}{2\pi}\right)^2\int\,dt[\p_t b_{c,\perp}-(\p_\perp b_0)_c]\frac{1}{\p_t^2+\omega_c^2}[\p_t b_{c,\perp}-(\p_\perp b_0)_c].\label{eq:Lag}
\end{align}
For the case of homogeneous $b(t,\vec x)=b(t)$ and with the radiation gauge condition $b^0=0$, the action simplifies to
\begin{align}
    \int\frac{d\omega}{2\pi}\left[\frac{m}{4\pi}i\omega\epsilon^{ij}+\frac{m}{4\pi\omega_B}\omega^2\delta^{ij}\right]A_H b_i(-\omega)b_j(\omega)-\frac{\alpha}{2\pi\varepsilon}\int\frac{d\omega}{2\pi}\frac{\omega^2}{\omega^2-\omega_c^2}b_{c,\perp}(-\omega) b_{c,\perp}(\omega),\label{eqsup:effectiveaction}
\end{align}
where $A_H $ is the area of the Hall bar and $\alpha = \frac{e^2}{4\pi}$. For the second contribution we note that, in the homogeneous case,
\begin{equation}
    b_{c,i}(t) = \int d^2x\phi_{k_c}|_{H}(\vec x)b_i(t)=\frac{\chi_cA_H }{\sqrt{V_c}}b_{i}(t).
\end{equation}
Thus the saddle-point equations following from \eqref{eqsup:effectiveaction} are
\begin{align}
    &M^{ij}(\omega)b_j(\omega)=0, &M^{ij}(\omega) = \left[\frac{m}{4\pi}i\omega\epsilon^{ij}+\frac{m}{4\pi\omega_B}\omega^2\delta^{ij}\right]A_H  -\frac{\alpha}{2\pi\varepsilon}\frac{\omega^2}{\omega^2-\omega_c^2}\frac{\chi_c^2A_H ^2}{V_c}\epsilon^{im}e_m\epsilon^{jn}e_n.
\end{align}
Nonzero solutions only appear at the frequencies which solve the characteristic equation $\det M(\omega)=0$,
\begin{equation}
    \omega^2-\frac{2\alpha\omega_B}{m}\frac{\chi_c^2A_H }{\varepsilon V_c}\frac{\omega^2}{\omega^2-\omega_c^2}-\omega_B^2=0,\label{eqsup:detomega}
\end{equation}
where we factored out the trivial root $\omega^2=0$. Since we are interested in the correction to the Kohn mode frequency $\omega\approx\omega_B\gg\omega_c$, we can simplify the denominator of the second term, giving
\begin{equation}
    \omega^2=\omega_B^2+\frac{2\alpha}{m L_{\rm eff}}\omega_B+O\left(\frac{\omega_c^2}{\omega_B^2}\right).\label{eq:kohnsup}
\end{equation}
Expanding \eqref{eq:kohnsup}, we find the leading correction to the Kohn mode frequency as
\begin{equation}
    \omega = \omega_B + \frac{\alpha}{m L_{\rm eff}}+O\left(\frac{\alpha^2}{m^2 L_{\rm eff}^2\omega_B^2}\right).
\end{equation}
For another solution of Eq.~(\ref{eqsup:detomega}) around the cavity frequency, such that  $\omega\approx\omega_c\ll\omega_B$, we have
\begin{equation}
    \omega=\omega_c-\frac{\alpha}{m L_{\rm eff}}\frac{\omega_c}{\omega_B}+O\left(\frac{\omega_c}{\omega_B},\frac{\alpha^2}{m^2 L_{\rm eff}^2\omega_B^2}\right).
\end{equation}

\section{III. Microscopic perspective}
Researchers in Ref.~\cite{Rokaj2023} have shown the conductivities of 2d free electron gas in a magnetic field can be calculated from the Kubo formula as follows~\cite{kubo1965}
\begin{align}
    \sigma^{ab}(\omega)&=-\frac{i}{\omega+i0^+}\left[\frac{e^2 n_e}{m_e}\delta^{ab}+\frac{\chi^{ab}(\omega)}{A_H }\right],\label{eq:sigma_ab}\\
    \chi^{ab}(\omega)&=\sum_{\mathbf{m}} \frac{\langle \mathbf{0}| J^a\left|\mathbf{m}\right\rangle\left\langle \mathbf{m}\right| J^b|\mathbf{0}\rangle}{\omega+\left(E_{\mathbf{0}}-E_{\mathbf{m}}\right)+\mathrm{i}0^+}-\left(\mathbf{0}\leftrightarrow \mathbf{m}\right),\label{eq:chi_ab}
\end{align}
where $n_e$ is the electron density, $A_H$ is the area of the Hall bar, $J^a$ is the total current operator, and $|\mathbf{m}\rangle$ is a general eigenstate of the Hamiltonian in the center of mass coordinate. In Landau gauge, this Hamiltonian for the wave functions with a plane-wave part $\exp(ikx)$ can be expressed as
\begin{align}
    H_\text{CM}&=\frac{(\vec{\pi}+e\sqrt{N}\vec{A}_c)^2}{2m_e}+\omega_c(a^\dagger a+\frac{1}{2}),\label{eq:Hcm}
\end{align}
where the $\pi^i$ satisfy $[\pi^x,\pi^y]=-ieB$ and $a$ ($a^\dagger$) is the annihilation (creation) operator of the cavity photons satisfying  $[\hat{a},\hat{a}^\dagger]=1$. 

For the DC conductances, previous results have been discussed only for a linear cavity such as $ \vec{A}_c=A\left(a+a^\dagger\right)\hat{e}_x$~\cite{Rokaj2023,Rokaj2024}. In fact, we can derive the DC conductances without specifying the polarization of the cavity. To show that, we first rewrite the current operators as
\begin{align}
 J^a&=\varepsilon^{ab}\frac{ i \sqrt{N}}{ B}[\pi^b,H_\text{CM}],
\end{align}
for the Hamiltonian Eq.~(\ref{eq:Hcm}) of center of mass, where $\varepsilon^{ab}$ is the anti-symmetric tensor with $\varepsilon^{xy}=-\varepsilon^{xy}=1$ and $\varepsilon^{xx}=\varepsilon^{yy}=0$. Then, we apply the new expressions of the current operators to Eq.~(\ref{eq:sigma_ab}) and obtain the AC Hall conductance
\begin{align}
    \sigma_{xy}(\omega) &=\frac{ i n_e}{B^2\omega }\sum_{\mathbf{m}}(E_{\mathbf{m}\mathbf{0}}^2-\omega^2+\omega^2)\bigg[\frac{\langle\mathbf{0}|\pi_y|\mathbf{m}\rangle\langle\mathbf{m}|\pi_x|\mathbf{0}\rangle}{\omega+E_{\mathbf{0}\mathbf{m}}}-\frac{\langle\mathbf{m}|\pi_x|\mathbf{0}\rangle\langle\mathbf{0}|\pi_y|\mathbf{m}\rangle}{\omega+E_{\mathbf{m}\mathbf{0}}}\bigg]\nonumber\\
    &=\frac{ e n_e}{B}\bigg\{i\langle\mathbf{0}|\frac{[\pi_x,\pi_y]}{eB}|\mathbf{0}\rangle+\sum_{\mathbf{m}}\frac{i\omega}{eB}\bigg[\frac{\langle\mathbf{0}|\pi_y|\mathbf{m}\rangle\langle\mathbf{m}|\pi_x|\mathbf{0}\rangle}{\omega+E_{\mathbf{0}\mathbf{m}}}-\frac{\langle\mathbf{m}|\pi_x|\mathbf{0}\rangle\langle\mathbf{0}|\pi_y|\mathbf{m}\rangle}{\omega+E_{\mathbf{m}\mathbf{0}}}\bigg]\bigg\},\label{eq:DC_xy}
\end{align}
and the longitudinal conductances 
\begin{align}
    \sigma_{aa}(\omega)&=-\frac{ i }{\omega  A_H}\frac{e^2 n_e}{m}-\frac{ i n_e}{B^2\omega }\sum_{b}\bigg\{|\varepsilon^{ab}|\langle\mathbf{0}|[\pi_b,[\pi_b,H]]|\mathbf{0}\rangle+\sum_{\mathbf{m}}\omega^2\frac{2E_{\mathbf{m}\mathbf{0}}}{\omega^2-E_{\mathbf{0}\mathbf{m}}^2}|\varepsilon^{ab}\langle\mathbf{0}|\pi_b|\mathbf{m}\rangle|^2\bigg\}\nonumber\\
    &=\frac{ e n_e}{B}\sum_{\mathbf{m},b} i \omega\frac{2E_{\mathbf{m}\mathbf{0}}}{E_{\mathbf{m}\mathbf{0}}^2-\omega^2}|\varepsilon^{ab}\langle\mathbf{0}|\frac{\pi_b}{\sqrt{eB}}|\mathbf{m}\rangle|^2.\label{eq:DC_aa}
\end{align}
Here, we define $E_{\mathbf{m}\mathbf{0}}=E_{\mathbf{m}}-E_{\mathbf{0}}$ for simplicity. For fully occupied Landau levels with filling ratio $\nu$, Eq.~(\ref{eq:DC_xy}) gives the quantized DC Hall conductances as $ e^2\nu/(2\pi)$ by using $n_e=eB\nu/h $ and $[\pi^y,\pi^x]=ieB$. Also, Eq.~(\ref{eq:DC_aa}) shows that the DC longitudinal conductances all vanish in the zero-frequency limit. Note that our derivation does not rely on a specific polarization of the cavity mode.

Using Eqs.~(\ref{eq:DC_xy}) and (\ref{eq:DC_aa}), we obtain the AC conductances for the $\hat{e}_y$-polarized cavity mode in the microscopic formulation as
\begin{align}
    \sigma^{xx}(\omega)&=\frac{e^2\nu}{2\pi}\left(\frac{i\omega\omega_B\cos^2\frac{\theta}{2}}{\omega_+^2-\omega^2}+\frac{i\omega\omega_B\sin^2\frac{\theta}{2}}{\omega_-^2-\omega^2}\right),\label{eq:exact_xx}\\
     \sigma^{yy}(\omega)&=\frac{e^2\nu}{2\pi}\left(\frac{i\omega}{\omega_B}\frac{\omega_+^2\cos^2\frac{\theta}{2}}{\omega_+^2-\omega^2}+\frac{i\omega}{\omega_B}\frac{\omega_-^2\sin^2\frac{\theta}{2}}{\omega_-^2-\omega^2}\right),\\
     \sigma^{xy}(\omega)&=\frac{e^2\nu}{2\pi}\left(1+\frac{\omega^2\cos{^2\frac{\theta}{2}}}{\omega_+^2-\omega^2}+\frac{\omega^2\sin{^2\frac{\theta}{2}}}{\omega_-^2-\omega^2}\right),
\end{align}
where $\omega_\pm=\omega_B-\omega_d(\cot\theta\mp\csc{\theta})$, $\tan\theta=2\omega_d/(\omega_B-\widetilde{\omega}_\text{c}^2\omega_B^{-1})$, with the modified cavity frequency $ \widetilde{\omega}^2_\text{c}=\omega_\text{c}^2+\omega_d^2$, and the diamagnetic frequency $\omega_d=\sqrt{e^2 N/({m_e \varepsilon V_c}}) = \sqrt{2\alpha\nu\omega_B/L_{\rm eff}}$,
where we recognize the characteristic energy scale $\omega_d^2/\omega_B=2\nu\alpha/ L_{ \rm eff}$ of the light-matter coupling in the cavity.  By substituting $\hat{e}_x$ to $\hat{e}_y$ in the AC conductances from Ref.~\cite{Rokaj2023}, one gets the same results. If we consider the regime where the $\omega_d\ll\omega_B$, the conductivities can be expanded to the second order of $\omega_d/\omega_B$
\begin{align}
     \sigma^{xx}(\omega)&\approx\frac{e^2\nu}{2\pi}\bigg[\frac{i \omega\,\omega_B}{\omega_B^2-\omega^2}+\frac{2\alpha \nu}{ L_\text{eff}}\frac{i \omega}{(1-\frac{\omega^2}{\omega_B^2})^2(\omega_c^2-\omega^2)}\bigg],\label{supeq:sigmaxxmicro}\\
      \sigma^{yy}(\omega)&\approx\frac{e^2\nu}{2\pi}\bigg[\frac{i \omega\,\omega_B}{\omega_B^2-\omega^2}+\frac{2\alpha \nu}{ L_\text{eff}}\frac{\omega^2}{\omega_B^2}\frac{i \omega}{ (1-\frac{\omega^2}{\omega_B^2})^2(\omega_c^2-\omega^2)}\bigg],\label{supeq:sigmayymicro}\\
   \sigma^{xy}(\omega)&\approx\frac{e^2\nu}{2\pi}\bigg[1+\frac{\omega^2}{\omega_B^2-\omega^2}+\frac{2\alpha \nu}{  L_\text{eff}}\frac{ \omega^2}{\omega_B(1-\frac{\omega^2}{\omega_B^2})^2(\omega_c^2-\omega^2)}.\bigg]
\end{align}
\begin{figure}
    \centering
    \includegraphics[width=0.9\linewidth]{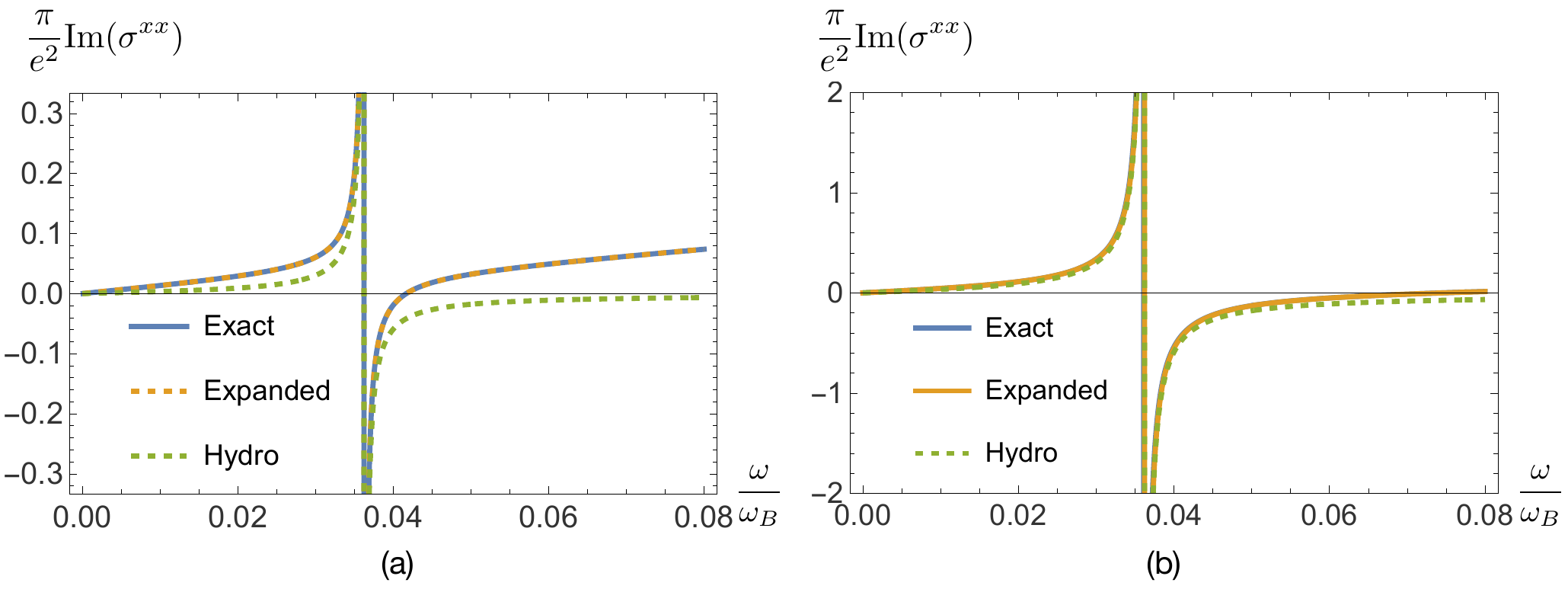}
    \caption{The AC conductivity $\sigma^{xx}$ for the cavity Hall state in the frequency region $\omega\in[0,0.08\omega_B]$ are computed using the exact results in Ref.~\cite{Rokaj2023} (Exact), Eq.(\ref{supeq:sigmaxxmicro}) (Expanded) and Eq.(\ref{eq:conduc_limit1}) (Hydro). In panel (a), the parameters are from recent experiments \cite{cavityQHE2022,Rokaj2024} for a $40\times200\mu\text{m}^2$ Hall bar with $\nu=1$, $\omega_B=24.3\text{THz}$ and in the cavity with frequency $\omega_c=0.88\text{THz}$, and cavity volume $V_0=2.7\times10^5\mu\text{m}^3$. In panel (b), the cavity volume is reduced to $0.1V_0$. We see that the expanded results by Eq.(\ref{supeq:sigmaxxmicro}) agree well with the full expression (Exact). Also, all results of  $\sigma^{xx}$ are consistent for a smaller cavitiy, in which case the Hydro contribution dominates.}
    \label{fig:ACcon}
\end{figure}
Comparing to the AC conductivities obtained from the hydrodynamics, the above results contain additional terms with a pole at $\omega_B$. Physically, this term is caused by the Landau level transitions. However, in the hydrodynamics, we have considered a regime where the cyclotron frequency $\omega_B$ is taken to be infinity in the Lagrangian Eq.~(\ref{eq:Lag}) and the effect from the Landau-level transition is neglected. In such a regime, we have the limits of the conductivities to the leading order of $\omega/\omega_B$ as
\begin{align}
       \sigma^{xx}(\omega)&\to \frac{(2\alpha \nu)^2}{ L_\text{eff}}\frac{i \omega}{(\omega_c^2-\omega^2)},\label{eq:conduc_limit1}\\
       \sigma^{yy}(\omega)&\to 0,\label{eq:conduc_limit2}\\
        \sigma^{xy}(\omega)&\to \frac{e^2\nu}{2\pi},\label{eq:conduc_limit3}
\end{align}
which agree with the results obtained by keeping only the topological terms in the hydrodynamic calculation. Note that the condition for this contribution to dominate the AC conductivity is
\begin{align}
\sigma^{xx}_B(\omega)=\frac{e^2\nu}{2\pi}\frac{i \omega\,\omega_B}{\omega_B^2-\omega^2}\ll \sigma^{xx}_c(\omega)=\frac{e^2\nu}{2\pi}\frac{2\alpha \nu}{ L_\text{eff}}\frac{i \omega}{(1-\frac{\omega^2}{\omega_B^2})^2(\omega_c^2-\omega^2)},
\end{align}
which, in the low-frequency regime $\omega\ll\omega_c\ll\omega_B$, yields 
\begin{align}
     \frac{2\alpha \nu}{ L_\text{eff}}\gg \frac{\omega_c^2}{\omega_B}.\label{eq:regime}
\end{align}
We find that some very recent experiments~\cite{cavityQHE2022,Rokaj2024} are actually not within this regime and the terms with the pole at $\omega_B$ cannot be neglected in the low-frequency region. Fixing the parameters of the Hall bar to the values in experiments~\cite{cavityQHE2022,Rokaj2024}, we find that the AC conductivities can roughly reach the limit of Eqs.~(\ref{eq:conduc_limit1}-\ref{eq:conduc_limit3}) if the effective length $L_\text{eff}$ is reduced. In Fig.~\ref{fig:ACcon}, we show the longitudinal conductivities in the function of $\omega/\omega_B$. The conductivities in (a) are for the $\nu=1$ Hall bar in the experiment~\cite{Rokaj2024}, where the cavity volume is $V_0=2.7\times10^5\mu\text{m}^3$. In (b), we set the cavity volume as $ 0.1V_0$ to make Eq.~(\ref{eq:regime}) satisfied.

\end{document}